\documentclass[preprint,showpacs,preprintnumbers,amsmath,amssymb]{revtex4}

\usepackage{graphicx}% Include figure files
\usepackage{dcolumn}% Align table columns on decimal point
\usepackage{bm}% bold math

\begin{document}

\preprint{}

\title{Multiscale Network Reduction Methodologies: Bistochastic and Disparity Filtering of Human Migration Flows between 
\newline 3,000+  U. S. Counties}

\author{Paul B. Slater}% 
\email{slater@kitp.ucsb.edu}
\affiliation{%
ISBER, University of California, Santa Barbara, CA 93106\\
}%
\date{\today}% It is always \today, today,
\newpage
\newpage
\begin{abstract}
To control for multiscale effects
in networks, one can, using the iterative proportional fitting (Sinkhorn-Knopp) procedure,  alternately scale the  row sums to equal 1 
and then the column sums, until sufficient convergence--under broad conditions--to a bistochastic (doubly-stochastic) table is ultimately  obtained. The dominant entries in the bistochasticized table can then be employed for network reduction 
purposes, using strong component hierarchical clustering. We illustrate various facets of this well-established, widely-applied two-stage algorithm with the $3, 107 \times 3, 107$ (asymmetric) 1995-2000 intercounty migration flow table for the United States. 
We compare the results obtained with ones using the disparity filter, for "extracting the "multiscale backbone of complex weighted networks", recently
put forth by Serrano, Bogu{\~n}{\'a} and Vespignani (SBV) ({\it Proc. Natl. Acad. Sci.} 106 [2009], 6483), upon which we have commented ({\it Proc. Natl. Acad. Sci.} 106 [2009], E66). In this specific migration-flow analysis, the performance of the bistochastic filter appears superior in two meaningful respects: (1) it requires far fewer links to complete a strongly-connected network backbone; and (2) it "belittles" small flows and nodes less--a principal desideratum of SBV--in the sense that the correlations of the nonzero raw flows are considerably weaker with the corresponding bistochastized links than with the significance levels yielded by the disparity filter. 
\end{abstract}

\pacs{Valid PACS 02.10.Ox, 02.10.Yn, 89.65.Cd, 89.75.Hc}
\keywords{networks; multiscale backbone;  bistochastic filter; disparity filter; seriation; transaction flows; hubs; clusters; matrix plots; internal migration; flows; U. S. intercounty migration; hierarchical cluster analysis; strong components; graph theory;  dendrograms; eigenanalyses; cosmopolitan areas; provincial areas;  functional regions; migration regions; socioeconomic networks;  functional regions;  functional regionalization; graph-theoretic clustering; network reduction}

\maketitle
\tableofcontents
\section{Introduction}
One of our principal objectives is to bring to the attention of the  recently-emerging  community of network scientists, a 
demonstratedly-insightful ("two-stage") methodology \cite{PBSPNAS}
developed and applied primarily in the period 1975-1985, largely in the substantive, socioeconomic contexts  of
internal migration flows between geographic subdivision of various nations of the world \cite{bellblake} and other forms of "transaction flows" 
\cite{savage,brams}. A second main goal is to employ this methodology anew, showing how it can be used to reveal the backbone of complex (in general) weighted, directed networks, while presenting a comparison of 
its performance with that of another "backbone-extraction" procedure (disparity filtering), more recently proposed  by Serrano, Bogu{\~n}{\'a} and Vespignani (SBV) \cite{SBV}. We apply certain  comparative 
(correlation and strong-connectedness) measures (sec.~\ref{Comparisons}), in this regard. 
However, we  limit our immediate conclusions as to the relative effectiveness of the two procedures for backbone-extraction 
to the specific 1995-2000 U.  S. intercounty migration data set that will serve as our main object of investigation.

In secs.~\ref{DisparityFilter} and \ref{BistochasticFilter}, we discuss the two multiscale reduction  methodologies (cf. 
\cite{Radicchi} \cite{Glattfelder})--disparity and bistochastic 
filtering.
We can, preliminarily,  contrast the two procedures in the following manner:
in the bistochastic approach, the square matrix of network flows is converted into a {\it single} bistochastic (doubly-stochastic) matrix, while in the disparity approach of SBV, in the general asymmetric case, it is transformed into a pair of row-stochastic and column-stochastic matrices. Both approaches, then, use their associated matrices for the construction of network backbones. In the bistochastic procedure, the entries of the associated matrix are themselves employed as linkage values, while in the disparity approach, the matrix entries are mapped to significance levels, which are then used (applying either $AND$ or $OR$ rules) for 
backbone construction. Both approaches are invariant under transposition of the original flow matrix, and/or multiplication of rows or columns by scaling factors. (This seems somewhat surprising, since one might expect the statistical significance levels [$\alpha$'s] in the disparity filter to change with sample size  (cf. \cite[p. 2]{mosteller}). Analyses involving either or both procedures should make explicit whether the presence of zero flows is considered to be due to structural or sampling considerations.)

We illustrate the use  and properties of the 
two-stage (bistochastic-filtering) algorithm, employing the large-scale example of migration between the more than three thousand county-level units of the United States 
(sec.~\ref{Data}) (cf. \cite{dirk1,dirk2} for analyses of U. S. intercounty flows of one-dollar bills, using the exceptional "Where's George?" database). 
In an analysis section (sec.~\ref{Comparisons}), we conduct studies of a similar nature to SBV, 
for comparative purposes (as called for in \cite{SBV2}). In a discussion section (sec.~\ref{RelatedIssues}) we comment on issues arising in our exchange of letters in the June 23, 2009 {\it Proceedings of the National Academy of Sciences} \cite{PBSPNAS,SBV2} with SBV, as well as  outline earlier, interesting results obtained using the two-stage algorithm. In sec.~\ref{Summary}, we present a summary of our findings here.

Our principal finding of a comparative nature here is that the bistochastic filter appears, in the migration example at hand, to outperform the disparity filter in, at least, two significant features: (1) the number (25,329) of bistochastic links needed to generate a strongly-connected backbone is far fewer than the number (lying  within the range 80,204 to 83,692) required by the disparity filter (using the $OR$ rule, preferred "because it ensures that small nodes in terms of strength are not belittle[d]" \cite[p. S3]{SBV}); and (2) the correlation of the 
logarithms of the 735,531 nonzero migration flows with the corresponding logarithms of bistochasticized values is considerably {\it weaker} than with the significance levels yielded by the disparity  filter (the same form of conclusion holding without taking the logarithms, with all the pertinent correlations, however, being somewhat weaker in nature)--thus, "belittling" small flows less, a 
principal desideratum of SBV.
\section{Disparity filter} \label{DisparityFilter}
A recent paper \cite{SBV} in the {\it Proceedings of the National Academy of Sciences} by Serrano, Bogu{\~n}{\'a} and Vespignani (SBV) entitled, "Extracting the multiscale backbone of complex weighted networks," has an abstract that reads--in part--as follows:
\begin{quote}
In recent
years, the study of an increasing number of large scale networks has highlighted
the statistical heterogeneity of their interaction pattern, with degree and weight
distributions which vary over many orders of magnitude. These features, along with
the large number of elements and links, make the extraction of the truly relevant
connections forming the network's backbone a very challenging problem. More
specifically, coarse-graining approaches and filtering techniques are at struggle
with the multiscale nature of large scale systems. Here we define a filtering
method that offers a practical procedure to extract the relevant connection
backbone in complex multiscale networks, preserving the edges that represent
statistical significant deviations with respect to a null model for the local
assignment of weights to edges. An important aspect of the method is that it does
not {\it belittle} (emphasis added) small-scale interactions and operates at all scales defined by the
weight distribution.
\end{quote}

The disparity filter that SBV advance, in the general weighted, directed case, takes  the form \cite[eqs. (8), (9) in SI]{SBV},
\begin{equation}
\alpha_{ij}^{in}=1 - (k^{in}-1) \int_{0}^{p_{ij}^{in}} (1-x)^{k^{in}-2} dx < \alpha,
\end{equation}
\begin{equation}
\alpha_{ij}^{out}=1 - (k^{out}-1) \int_{0}^{p_{ij}^{out}} (1-x)^{k^{out}-2} dx < \alpha.
\end{equation}
Here $\alpha$ is a preassigned significance level, $k^{in}$ and $k^{out}$ denote the in-degree and out-degree "of the node to which the directed link under consideration is attached", and $p_{ij}^{in}$ and $p_{ij}^{out}$ indicate the associated transition probabilities (normalized weights). One can employ an $AND$ rule or an $OR$  rule on the pair $\{\alpha_{ij}^{in},\alpha_{ij}^{out}\}$ for testing the significance of the $ij$-link, and thus deciding whether or not to admit it into the backbone. 
(SBV expressed a preference for the application of the $OR$ rule  "because it ensures that small nodes in terms of strength are not belittle[d]" \cite[SI, p. 3]{SBV}.) For comparative purposes, SBV also apply a global threshold filter, which destroys "the multiscale nature of more realistic networks where weights are locally correlated on edges incident to the same node and nontrivially coupled to topology" \cite[p. 6483]{SBV}.

The null hypothesis underlying the use of the disparity filter is that the 
incoming (outgoing) connections of a node are produced by a uniform
random assignment. In \cite{SBV3} the disparity filter has been applied 
by SBV to the world trade web to find "dominant trade channels" (cf. \cite{savage,Goodman,brams,schwarz,TradeItaliani}). While the disparity filter is based on {\it significance-testing}, the bistochastic filter with which it is to be compared here, and discussed immediately next, relies upon an {\it estimation} procedure to generate {\it measures of association} \cite{mosteller}.
\subsection{The GloSS filter of Radicchi, Ramasco and Fortunato}
We note that in regard to the disparity filter, Radicchi, Ramasco and Fortunato recently stressed "that treating vertices independently of each other is risky when the objects of the investigation are the edges, which join pairs of vertices" \cite[p. 2]{Radicchi}. As an alternative, they proposed "a weight filtering  technique based on a global null model (GloSS filter), keeping both the weight distribution and the full topological structure of the network". Their null model consists of a graph in which 
"the connections of the original network are locked, while weights are assigned to edges by randomly extracting values from the observed weight distribution". They conclude--from a number of specific analyses--that the "significance of the edges is not so strongly correlated with their weights like for other techniques". Of course, at some point, it would also be of interest to incorporate this filtering procedure into the general comparisons reported below.
\section{Bistochastic filter} \label{BistochasticFilter}
The underlying (network reduction) motivations of SBV in devising the disparity filtering methodology appear to be somewhat similar to those for a certain two-stage  algorithm the use of which was first reported in 1975 \cite{russia,turkish,metron}. Over the succeeding decade, this methodology was widely applied to internal migration flows between the geographic subdivisions of numerous nations and other forms
of "transaction flows" (see the extensive bibiliography in \cite{SlaterDendrogram}). Many of these applications were collected in 
the 1984 research 
institute survey monograph \cite{tree}. (In a review of 
\cite{tree}, R. C. Dubes wrote that the two-stage methodology "might very well be the most successful application of cluster analysis" \cite{dubes}.)

SBV remark that "Reduction schemes can be divided into two main
categories: coarse-graining and filtering/pruning". The two-stage procedure can readily be seen to fulfill a role in both categories.
\subsection{First stage of the two-stage algorithm}
In the first stage (iterative proportional fitting procedure [IPFP] 
\cite{fienberg}),
the rows and columns of the 
table of flows ($f_{ij}$)
are alternately  ("biproportionally" \cite{bacharach}) scaled 
to sum to a fixed number (say 1). Under broad conditions--to be discussed 
shortly--convergence occurs to a ``doubly-stochastic" 
(bistochastic) 
table, with row and column sums all 
{\it simultaneously} equal to 1 \cite{mosteller,louck,CSBZ,unistochastic,
romney,wong}. 
The purpose of the scaling is to 
remove overall (marginal) effects of size, and focus on relative, 
interaction  effects. 
Nevertheless, the {\it cross-product ratios} 
({\it relative odds}), $\frac{f_{ij} f_{kl}}{f_{il} f_{kj}}$, 
measures of association \cite{mosteller}, 
are left {\it invariant}. 
Additionally, the entries of the
doubly-stochastic table provide 
{\it maximum  entropy} estimates of the original
flows, given the row and column constraints \cite{eriksson,macgill}.
Further, the $ij$-entry of the bistochastic table can be written
as the product of the raw flow $f_{ij}$ and a row multiplier ($r_{i}$) 
and a column multiplier ($c_{j}$). 

For large {\it sparse} 
flow tables, only the nonzero entries, together with their
row and column coordinates are needed for the IPFP. Row and column (biproportional) 
multipliers ($r_{i}, c_{j}$) can be iteratively computed by sequentially accessing the nonzero
cells \cite{parlett}. If the table is ``critically sparse'', various convergence difficulties may occur. Nonzero entries that are ``unsupported''--that is, not part of a set of $N$ nonzero entries, no two in the same row and 
column-- may converge to zero and/or the 
biproportional multipliers may not converge \cite[p. 19]{tree} \cite{sinkhorn} \cite[p. 171]{mirsky}.
The ``first strongly polynomial-time algorithm for matrix scaling'' 
was reported in \cite{linial}.

{\it Smoothing} procedures
could be used to modify the zero-nonzero structure 
of a flow table--treating the zeros as due to {\it sampling}, rather than {\it structural} effects--particularly 
if the table is 
critically sparse \cite{simonoff,boundary}. If one takes the 
second power of 
a doubly-stochastic matrix, one obtains another such 
matrix--of predicted {\it two}-step movements--but 
smoother in character. One might also 
consider standardizing the {\it i}th row [column] sum 
to be proportional to the number of non-zero entries in the 
{\it i}th row [column]--although we found considerable numerical 
difficulties when attempting this, using the methodology developed in \cite{linial}-- for the 1995-2000 U. S. intercounty
migration table analyzed below. Another procedure--in line with
the Google page-ranking [``teleporting random walk''] 
procedure \cite{brinpage,langville}, that has been much studied and 
emulated--is to take some convex combination of the doubly-stochastic
table and the $N \times N$ table all the off-diagonal entries of which
are equal to $\frac{1}{N-1}$.
W. D. Smith has suggested in an apparently unpublished 2005 paper 
(locatable through the www.scholar.google.com website), entitled
"Sinkhorn ratings, and new strongly polynomial time algorithms for Sinkhorn 
balancing, Perron eigenvectors, and Markov chains", that Laplace's rule
from Bayesian statistics suggests the use of an "add one" idea that the value 1 should be added to each cell of the table (thus, obviating any
convergence problems). 

\subsection{Second stage of the two-stage algorithm}
In the second stage of the two-stage 
procedure, the doubly-stochastic matrix 
is converted to a series of {\it directed} 
(0,1) graphs (digraphs), by applying thresholds to its entries. 
As the thresholds   are 
progressively lowered, larger and larger {\it strong components} 
(a directed path existing from any member of a component 
to any other) of the resulting 
graphs are found. This process 
(a simple variant of well-known single-linkage [nearest-neighbor or min]
clustering \cite{gower1}) can be represented by the familiar dendrogram 
or tree diagram used in 
hierarchical cluster analysis and cladistics/phylogeny (cf. \cite{ozawa,hubert,tumminello2}). (The ``CLASSIC'' methodology proposed by Ozawa--though couched in
rather different terminology--appears to be fully equivalent. 
Ozawa found the procedure to be 
useful in ``the detection of gestalt clusters'' 
\cite{ozawa}.)
\subsection{Computer implementation}
A FORTRAN implementation of the two-stage process was given in 
\cite{leusmann} (and extensively applied in \cite{gawryszewski}), as well as a realization 
in the SAS (Statistical Analysis System) 
framework \cite{chilko}. Subsequently, 
the noted computer scientist (1982 Nevanlinna medalist) R. E. Tarjan 
\cite{schwartz} devised an $O(M (\log{N})^2)$
algorithm \cite{tarjan} for strong component hierarchical clustering, 
and, then, a further improved $O(M (\log{N}))$ 
method \cite{tarjan2}, 
where $N$ is the number of nodes and $M$ the number of edges of 
a directed graph. (These substantially improved upon 
the earlier works \cite{leusmann,chilko}, 
which 
required the 
computations of {\it transitive closures} of graphs--in terms of 
which the analysis of Ozawa \cite{ozawa} is phrased--and were 
$O(M N)$ in nature.) A FORTRAN coding--involving 
linked lists--of the improved Tarjan 
algorithm \cite{tarjan2} was presented
in \cite{tarjanslater}, and applied in the 
1965-70 US intercounty study \cite{county}. 
If the graph-theoretic (0,1)-structure of a network under study 
is {\it not} strongly connected 
\cite{hartfiel}, {\it independent} two-stage analyses of 
the subsystems of the network would be appropriate. 
(So, it is interesting to note that there does exist some form of mathematical relationship between the two apparently quite distinct stages of the two-stage algorithm.)
\subsection{Further background}
In the recent spate of activity and interest in the science of networks over the previous decade or so,
the two-stage algorithm has been little applied nor analyzed, it seems. Neither, does it appear to have been re-invented. (Perhaps of relevance in this regard is that a number of applications of the two-stage algorithm and associated comments/discussions appeared in the journal {\it IEEE Transactions on Systems, Man and Cybernetics}, and that much 
scientific attention has  shifted over the years from "systems analysis" to "network analysis".)

In 2008, our attention was redrawn--after a long hiatus--to this general area of research, by the preparation of a review \cite{siegfriedReview} of "A Beautiful Mind: John Nash, game theory, and the modern quest for a code of nature"
\cite{siegfried}. We, then, posted papers in which we sought to bring the interesting properties and many applications of the two-stage algorithm to the attention of network analysts \cite{discernment,SlaterDendrogram}. 
In particular, we have had a letter published \cite{PBSPNAS}, commenting on the recent SBV paper \cite{SBV}, elaborated upon above, to which SBV have responded \cite{SBV2}. The issues arising in this interchange form the basis for this study.

\section{Two-stage analysis of U. S. intercounty migration table} \label{Data}
\subsection{Matrix of Intercounty Flows}
Based upon a question as to responders' 1995 residences posed in the 2000 United States Decennial Census, one can construct a 
square (origin-destination) matrix of 1995-2000 migration flows ($m_{ij}$) 
between
3,107 county-level units of the nation. 
Many nations in their censuses ask similar questions as to previous residence. 
The results are often reported in the form of "internal migration 
tables"--which have  served as a basis for most of  the two-stage 
multinational analyses reported in \cite{tree}. For a very comprehensive multi-author review of the issues arising in the analysis and interpretation of data of this nature, entitled: "Cross-National Comparison of Internal Migration: Issues and Measures", see 
\cite{bellblake}.)
\subsection{Matrix plot of raw unadjusted flows}
\begin{figure}
\includegraphics{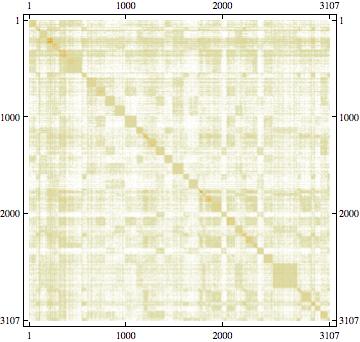}
\caption{\label{fig:matrixplotraw}Unadjusted (raw) 1995-2000 intercounty
U. S. migration table. The states are first ordered alphabetically, and then the counties alphabetically within the states. The large square near the end of the
diagonal corresponds to the state with the most (254) counties, Texas, 
while Georgia, with 159 counties, is located near the beginning. Counties 1000 and 2000--which Mathematica chooses to 
locate--are Boyd County, Kentucky and Dunn County, North Dakota, respectfully.}
\end{figure}

In Fig.~\ref{fig:matrixplotraw}, 
we show a matrix plot
of this (raw data) 
table. (In the absence of any further relevant information, we set to zero
the diagonal entries--which conceptually 
might correspond either to the number of people who actually moved within 
the county or who simply stayed within it  (cf. \cite{deMontis}).) 
In the principal, admininstrative  
sorting of the rows/columns of the table, the fifty states are ordered alphabetically, while, secondarily,  
within the states, their constituent counties are ordered also alphabetically. 

We immediately discern a clear clustering along the diagonal in 
Fig.~\ref{fig:matrixplotraw}, indicative of the obvious proposition that migrants have a proclivity to move intrastate-wise, both
for simple distance and state loyalty/ties/allegiance considerations.
However, the alphabetical ordering by states is certainly 
highly fortuitous in
character, and we observe relatively heavy migration far removed from
the diagonal (say for the physically contiguous, but alphabetically 
non-proximate pairs [California, Oregon] and [Lousiana, Texas].)
(Historically, the 
design and layout of counties differ considerably--somewhat 
unfortunately from a geographic-theoretic 
point of view--between states, and we note
that Texas has the most counties, 254, and appears as a large square
far down the diagonal in Fig.~\ref{fig:matrixplotraw}, while 
the state of Georgia, 
with the second most counties, 159, 
is also apparent near the upper left corner. In these and subsequent matrix plots, zero values are displayed as white, with negative values tending to be bluish and positive values reddish.)

\subsubsection{Multiscale character of U. S. intercounty migration flows}
\begin{figure}
\includegraphics{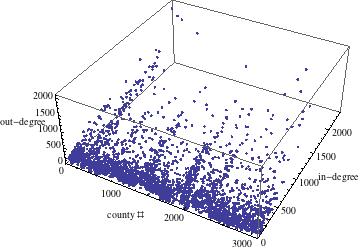}
\caption{\label{fig:degrees}Joint plot of administrative county number and in- and out-degrees of the 3,107 counties, that is, the number of non-zero entries in rows and columns 
of the intercounty migration table. The ordering of counties is the same as in the first figure.}
\end{figure}

\begin{figure}
\includegraphics{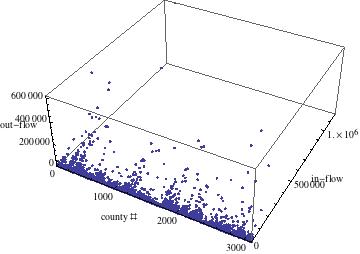}
\caption{\label{fig:flows} Joint plot of adminstrative county number and the 1995-2000 in-migrant and out-migrant totals for each of the 3,107 counties.  The largest in-flows are to Los Angeles, California, Chicago [Cook County], Illinois and Houston [Harris County], Texas.  The ordering of counties is the same as in the first two figures.}
\end{figure}

In Fig.~\ref{fig:degrees}, we jointly plot the county number (1 to 3,107) in the administrative order employed in the two previous figures, along  with the out-degrees and the in-degrees (that is, the number of counties receiving and sending migrants 
to and from a specific county), and in Fig.~\ref{fig:flows}, we analogously employ the total
in- and out-migrants for each county. 
(The correlation between in- and out-data for Fig.~\ref{fig:degrees} is  0.96523 and 
0.92336 for Fig.~\ref{fig:flows}. The largest in-degrees are for Los Angeles [2,371], Maricopa [Phoenix, 2,259] and San Diego [2,243] Counties, respectively, while the largest out-degrees are for Maricopa [2,012], San Diego [1,853] and Los Angeles [1,587], respectively. In the administrative numbering system, Los Angeles is county $\#$203, Maricopa, $\#$102 and San Diego, $\#$2243.)
\subsection{First stage--bistochasticization of migration table}
\begin{figure}
\includegraphics{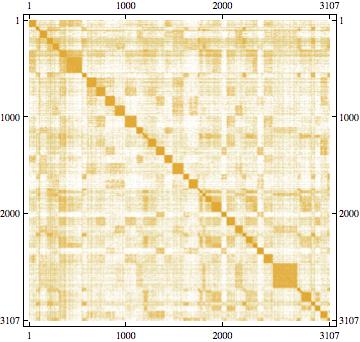}
\caption{\label{fig:matrixplotds}Doubly-stochastic form
of the 1995-2000 intercounty
U. S. migration table. Fuzziness in Fig.~\ref{fig:matrixplotraw} is greatly reduced. The correlation between the 735,531 non-zero entries of both these
forms of the table is 0.15712, which can be increased to 0.27905 by taking the logs of the 
raw flows.}
\end{figure}

Counties vary widely in their number of in- and out-migrants (Fig.~\ref{fig:flows}). To control for 
this (marginal/multiscale) 
effect, one may biproportionally/iteratively adjust the row and 
column sums (the "Sinkhorn-Knopp algorithm" \cite{knight}) so that all $6,214 = 2 \times 3,107$ such sums converge to be equal (say to 1). 
(This algorithm provides the basis of a measure--alternative to the PageRank employed by Google--of web page significance \cite{knight}.) In 
Fig.~\ref{fig:matrixplotds}, we show the $3,107 \times 3,107$ intercounty migration 
table after such a bistochasticization (double-standardization).
Clearly, the underlying definition/delimitation of blocks has been heightened 
by this transformation.
The purpose of the scaling is to 
remove overall effects of size (which certainly may be of
interest in themselves [Fig.~\ref{fig:flows}]), and focus on the usually more intricate relative, 
interaction  effects. 
Nevertheless, the {\it cross-product ratios} 
({\it relative odds}), $\frac{m_{ij} m_{kl}}{m_{il} m_{kj}}$, 
measures of association, are left {\it invariant} in the process. 
Additionally, interestingly, the entries of the
doubly-stochastic table provide 
{\it maximum  entropy} estimates of the original
flows, given the constraints on the row and column sums 
\cite{eriksson,macgill}. So, this corresponds to an idealized situation in which all counties were constrained to emit and receive the same numbers of migrants.
\subsubsection{Eigenanalyses of bistochastic table}
\begin{figure}
\includegraphics{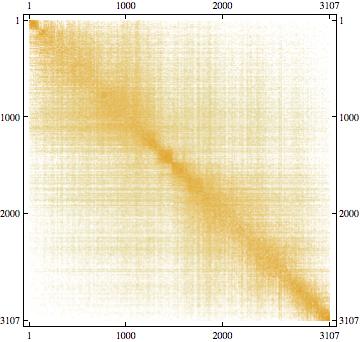}
\caption{\label{fig:subdominant}Doubly-stochastic matrix 
(Fig.~\ref{fig:matrixplotds}) {\it reordered} on the basis of its subdominant left eigenvector. The first
72 counties in the ordering are {\it all} from Georgia (mostly lying in a 
[``Upper Coastal Plain''] band from
the southwest corner of the state [Seminole County] 
to its north central boundary [Franklin, Hart, Elbert and Lincoln Counties]), 
and the last 110, all from the Great Plains states of 
North Dakota (45), South Dakota (50) and (north central) Nebraska (15). County 1000 is
Bucks County, Pennsylvania and 2000, Lubbock County, Texas.}
\end{figure}

\begin{figure}
\includegraphics{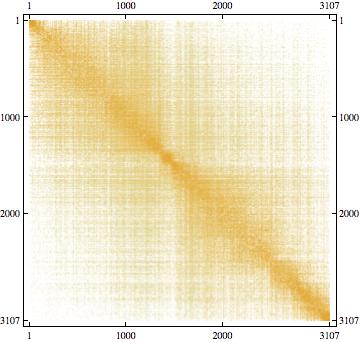}
\caption{\label{fig:subdominantTranspose}Doubly-stochastic matrix (Fig.~\ref{fig:matrixplotds}) reordered on the basis of its subdominant right eigenvector. Rather similarly to Fig.~\ref{fig:subdominant},
the first 74 counties in the ordering are all from Georgia, and the last 
181, all from North Dakota, South Dakota, Nebraska and \mbox{Min}nesota. County 1000 is Washington County, Louisiana and 2000, 
Adair County, Oklahoma.}
\end{figure}

The dominant left and right eigenvectors (corresponding to the eigenvalue 1) 
of the doubly-standardized table
are simply {\it uniform} vectors. The subdominant (left and right) eigenvectors
(corresponding to a {\it real} 
eigenvalue of 0.90625) are of interest 
\cite{meila}. (The correlation between these 
two eigenvectors is high, 0.97119. 
The third largest eigenvalue is real also, 0.86878, while the fourth
is slightly complex in nature, $0.84562 + 0.0009063 i$. The vector 
of 3,107 eigenvalues has length 12.6472.)
We {\it reorder} or {\it seriate} Fig.~\ref{fig:matrixplotds} on the basis of the left (in-migration) 
eigenvector and obtain Fig.~\ref{fig:subdominant}, 
and  on the basis of the right (out-migration) 
eigenvector and obtain Fig.~\ref{fig:subdominantTranspose}.
Now we see diminished clustering far from the diagonal. Further, both
of these figures suggest the division of the nation into basically two
large clusters.
\subsection{Second stage-strong component hierarchical clustering (SCHC)}
Further, reordering on the basis of the 
(38-page-long, 3,107-county) 
dendrogram (\cite[Supplement]{SlaterDendrogram} \cite{epaps}) generated by 
the strong component hierarchical clustering (the directed-graph 
analogue of the single-linkage method) \cite{japan,tree,france,science,partial,SEAS,qq,tarjan,tarjan2,ozawa} 
of the bistochasticized table, we obtain
Fig.~\ref{fig:ReOrdered1}. 
The correlation between the ordering used in Fig. 7 and the admininstrative ordering used in
Figs. 1 and 2 is 0.037352, and between the ordering used in Fig. 7 and the orderings used in Figs. 5 and 6, respectively, even lower,
0.004015 and 0.009995. (The corresponding correlations between the
administrative ordering and the seriations employed in Figs. 5 and 6 are 0.057925 and 0.075508. Correlations greater in 
absolute value 
than 0.035307 are significant at the $95\%$ level, 0.040065 at the $97.5\%$ level, and 0.045826 at the $99\%$ level.)
\subsubsection{Cosmopolitan or hub-like behavior of counties}
The dominant feature of Fig.~\ref{fig:ReOrdered1} is that 
the counties now listed at the beginning 
in the reordering--and, in general, the last to be absorbed in the
agglomerative clustering process--are ``cosmopolitan'' or 
``hub-like''. They tend to receive and send migrants across the nation, while those nearer to the end in 
the reordering tend to be more provincial or limited
in their breadth of interactions \cite{france}.
(A prototypical example of a hub-like internal migration area is Paris 
\cite{france,winchester}. In analytically parallel studies of interjournal citations 
\cite{science,rosvall,bollen}, one might anticipate that the broad journals, {\it Science}, {\it Nature} and the {\it Proceedings of the National Academy of Sciences} might fulfill analogous roles.) This appears to be an interesting feature of the two-stage algorithm {\it specific} 
to it.
\subsubsection{Ultrametric (hierarchical) fit to the bistochastic and residuals therefrom} 
The {\it ultrametric} fit to this reordered bistochasticized table 
provided by the strong component hierarchical clustering 
\cite{japan,tree,france,science,partial,SEAS,qq,tarjan,tarjan2,ozawa} 
is given in Fig.~\ref{fig:ultrametricFitReordered}, and the residuals 
(predominantly negative) from the hierarchical fit in 
Fig.~\ref{fig:residuals}. 
(These latter two figures, both in their  own ways, further 
reflect this cosmopolitan-provincial dichotomy between the 
U. S. counties.)
Let us also indicate that the two-stage methodology can be well-viewed well as  BOTH a filtering AND coarse-graining procedure. This serves to cast light and understanding upon the predominantly largely negative residuals from the ultrametric fit to the doubly-stochastic table. (Usually, in fits of statistical models, one finds some rough balance between negative and positive residuals.)
\subsection{Possible effect of non-zero {\it intra}county diagonal entries} \label{remarks}
Much earlier \cite{county,qq} than this current paper, we had also studied (but without the aid of the 
more recently-developed computerized matrix plots used above) bistochasticized forms of the 1965-70 U. S. intercounty
migration table with strong component hierarchical clustering 
\cite{japan,tree,france,science,partial,SEAS,qq,tarjan,tarjan2,ozawa},
{\it both} with zero and non-zero (corresponding to intracounty 
movements) diagonal entries. Counties with large physical areas tend
to absorb more of their own migrants, and thus exhibit larger
diagonal bistochasticized entries and smaller off-diagonal entries in the non-zero-diagonal analysis,
making them link at weaker levels in the dendrogram generated in the zero-diagonal analysis (cf. \cite{deMontis}). 

Journals with high self-citations would be expected to behave analogously
in journal citation-matrix analyses \cite{science,rosvall,loet}. 
(In the application of our two-stage bistochasticization and strong component
hierarchical clustering procedure to the 1967-75 interjournal citations between 
twenty-two mathematical journals, the {\it Proceedings of the American
Mathematical Society} was found to function in a 
particularly broad, cosmopolitan 
manner in a zero-diagonal analysis \cite{science}, while {\it Advances in 
Mathematics} played an analogous role when diagonal entries were taken into  account.)

\begin{figure}
\includegraphics{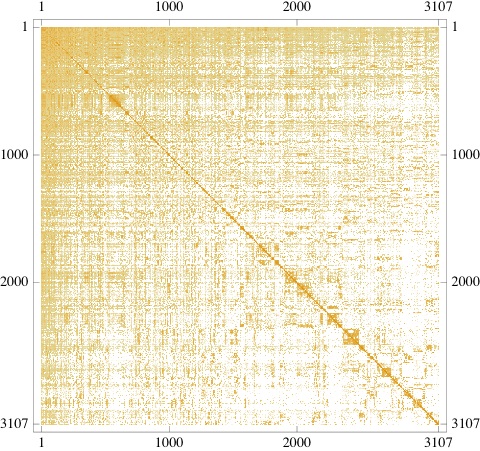}
\caption{\label{fig:ReOrdered1}Doubly-stochastic matrix 
(Fig.~\ref{fig:matrixplotds}) reordered on the basis of its strong component
hierarchical clustering. The first twelve (``cosmopolitan'')
counties in the seriation are
all from the ``Sunbelt'' states of Florida (5 counties,  a 
well-defined cluster of 
four of them
being equivalent to 
the Tampa-St. Petersburg-Clearwater Metropolitan Statistical 
Area), 
Arizona (2), (southern) California (3), Nevada (Las Vegas) (1)
and Texas (Dallas) 
(1). The last 35 (``provincial'') ones--lie 
principally in the ``Black Belt'', stretching  through the Deep South 
states of Mississippi (5),
Alabama (24), Georgia (4) and (Panhandle) Florida (2). County 1000 is Carroll County, 
Indiana and 2000, Warren County, New Jersey.}
\end{figure}
\begin{figure}
\includegraphics{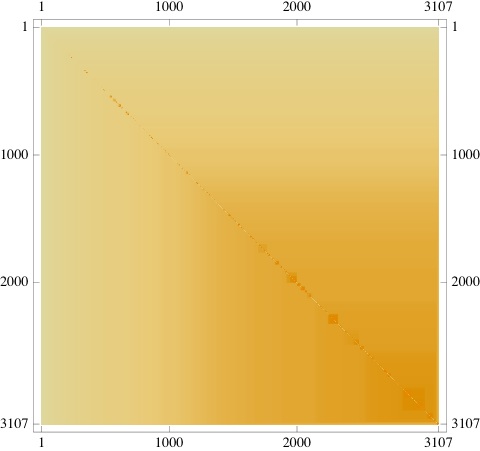}
\caption{\label{fig:ultrametricFitReordered}Ultrametric (strong component 
hierarchical 
clustering [SCHC]) fit to the doubly-stochastic matrix Fig.~\ref{fig:ReOrdered1}. 
The fits tend to be higher in the lower right-hand corner, corresponding
to the more ``provincial'' (including ``Black Belt'') counties.}
\end{figure}
\begin{figure}
\includegraphics{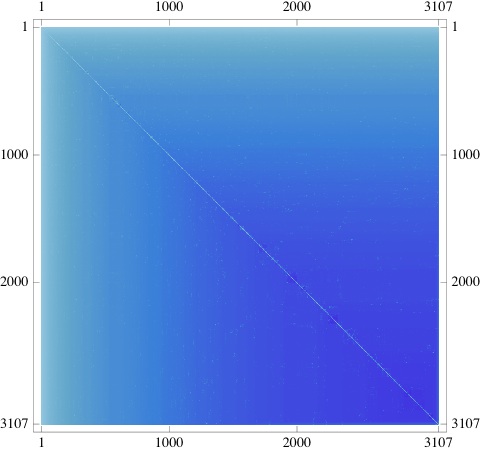}
\caption{\label{fig:residuals}Residuals (predominantly negative) of the 
ultrametric fit (Fig.~\ref{fig:ultrametricFitReordered}) to the doubly-
stochastic matrix (Fig.~\ref{fig:ReOrdered1}). The residuals are most
negative in the lower right-hand corner, where the fits 
provided by the strong component hierarchical clustering [SCHC] were greatest om value.}
\end{figure}
\section{Comparisons of bistochastic and disparity filters} \label{Comparisons}
In their response \cite{SBV2} to the letter \cite{PBSPNAS} commenting on their article
\cite{SBV}, 
Serrano, Bogu{\~n}{\'a} and Vespignani (SBV) have called for 
"an in-depth analysis of Slater's [two-stage] technique on a set of standard multiscale networks and a thorough comparison of the results with respect to ours and other methods, as we have done in our paper [\cite{SBV}]". Of course, this is a most appropriate proposal, which we now pursue in depth here.

The SBV methodology appears capable of producing a hierarchy of nodes, so direct comparisons in this regard should be possible \cite{Costa}. 
Additionally, we can choose to take as an obvious candidate for the multiscale backbone, the 25,329 links required in our intercounty migration 
two-stage analysis above to complete the strong 
component hierarchical clustering (SCHC) \cite{SlaterDendrogram}. (It would be of interest to 
overlay this backbone--both in raw and bistochastic form--on a county map of the United States [cf. \cite[Figs. 1-4]{SlaterDendrogram}]. If we, in a {\it global} filtering process, apply the SCHC to the largest raw migration flows, rather than to their bistochastic counterparts, it requires more than half-a-million, as opposed to 25,329 links, to complete the process.)

This can be compared with backbones generated by the techniques of SBV. 
(In \cite[p. 4]{Radicchi}, Radicchi, Ramasco and Fortunato wrote that the test of "how many edges are needed to form a connected graph \ldots 
has been suggested in" \cite{SBV}. In an e-mail clarification of this point, Fortunato wrote "What we referred to was Fig. 1 of the paper, in which they show how many nodes
remain in the largest connected component after removing (or adding, if you wish) 
some fraction of the edges according to their method".)
Of course, by choosing thresholds one can truncate links in the SCHC backbone with smaller bistochastic values 
to include precisely any specific number of links one {\it a priori} desires in the backbone ultimately selected.
\subsubsection{Correlations}
Let us here note that the correlation between the 735,531 nonzero bistochastic values of 
the $3,107 \times 3,107$ intercounty migration table and the corresponding signficance levels ($\alpha_{ij}$) of SBV is -0.33183, using an $OR$ rule, that is taking as the second variable $\mbox{Min}[\alpha_{ij}^{in},\alpha_{ij}^{out}]$ and -0.42094, with the use of an $AND$ rule, that is taking $\mbox{Max}[\alpha_{ij}^{in},\alpha_{ij}^{out}]$. Of course, we expect the correlations to be negative, since {\it smaller} 
$\alpha$'s  indicate {\it greater} significance. We can strengthen the two correlations to -0.58630 and -0.64089, respectively, by using the logarithms of the bistochastic values, rather than the bistochastic values themselves.

The (notably strong) correlations between the logarithms of the 735,531 nonzero raw (unadjusted) flows 
and the corresponding values of $\mbox{Min}[\alpha_{ij}^{in},\alpha_{ij}^{out}]$ 
is  -0.88681 and with $\mbox{Max}[\alpha_{ij}^{in},\alpha_{ij}^{out}]$, -0.84975. Thus, large raw flows certainly tend to be highly significant 
in the disparity filter model.
\subsubsection{Cumulative plots}
\begin{figure}
\includegraphics{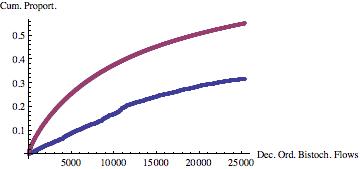}
\caption{\label{fig:SBVtype}Cumulative proportions of decreasingly ordered  bistochastic flows (higher red curve) and corresponding (non-ordered) raw flows (lower blue curve). The largest 25,329 bistochastic flows are needed to complete the strong component hierarchical clustering and can be considered to form the multiscale backbone of the network.}
\end{figure}

In the spirit of the analysis of SBV \cite{SBV}, and in response to their 
call \cite{SBV2} for further testing, we present 
Fig.~\ref{fig:SBVtype}. On the vertical axis--as a measure of total explanation--we plot the cumulative proportion (reaching 0.55029) of bistochastic flows (red curve)  and the cumulative proportion (reaching 0.31626) of the corresponding raw 
(unadjusted) migration flows (blue) curve, as functions of the 
decreasingly ordered (from 1 to 0.022542) 
largest 25,329 links in the bistochastic table. 

Let us note also that the correlation between
the largest 25,329 bistochastic values and the correponding raw [flow] values
is 0.045190, while the analogous correlation for the 735,531 non-zero
raw and bistochastic flows is larger, 0.15712. These can be increased to 0.26249 and 0.27905, respectively, by using the logarithms of the raw flows, and further still to 0.31837 and  0.40843, respectively, by taking the logs of both the raw and bistochastic variables. {\it Thus, large bistochastic values do exhibit a tendency to be associated with large raw values--but not as much as do the disparity filter significance levels}.

\begin{figure}
\includegraphics{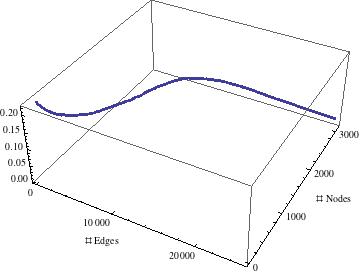}
\caption{\label{fig:SBVtype2}Evolution of the strong component hierarchical clustering (SCHC) process. As edges with decreasing bistochstic values 
($z$-axis) are introduced into the initially edge-less digraph, previously isolated nodes are incorporated into the 
multiscale backbone of the migration network, until eventually [with the 25,329th largest link] all nodes are joined together.}
\end{figure}

Further, in Fig.~\ref{fig:SBVtype2}, we show the evolution of the SCHC agglomerative clustering process. As edges associated with smaller bistochastic values
are introduced into the initially edge-less digraph, more previously isolated nodes are incorporated into nontrivial strong components, until with the 25,329th edge, associated with a bistochastic value of 
0.022542 (for the flow from Indian River County, Florida to Brevard County, Florida), all 3,107 nodes (counties) are joined together to complete 
the clustering process (as well as forming a candidate for the 
multiscale backbone of the migration network).

\begin{figure}
\includegraphics{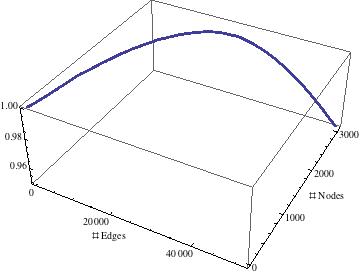}
\caption{\label{fig:SBVtype2disparity}Evolution of the disparity filter backbone of Serrano, Bogu{\~n}{\'a} and Vespignani \cite{SBV}, using $OR$ rule. As edges with decreasing values 
of $1 - \mbox{Min}[\alpha_{ij}^{in},\alpha_{ij}^{out}]$ (plotted along the $z$-axis) are introduced into the initially edge-less digraph, previously isolated nodes are incorporated into nontrivial strong components forming the multiscale backbone of the migration network. 
For each of 
the 51,900 links employed, 
$\mbox{Min}[\alpha_{ij}^{in},\alpha_{ij}^{out}] < 0.05$.
The use of these links still leaves 15 individually isolated nodes (counties) (six neighboring ones from north central Nebraska), unincorporated into the backbone.}
\end{figure}

We have constructed a comparable figure to Fig.~\ref{fig:SBVtype2} based on the disparity filter of SBV \cite[eq. (8), SI]{SBV}, 
using an $OR$ rule on the 
pair of in-flow and out-flow 
significance levels, $\alpha_{ij}^{in}, \alpha_{ij}^{out}$. (We subtract the results 
from 1, in order to more directly graphically compare results based on the bistochastic values.)
In Fig.~\ref{fig:SBVtype2disparity}, we show the evolution of the backbone as the significance level is raised.
It would be of interest to jointly plot the data in  Figs.~\ref{fig:SBVtype2disparity} and \ref{fig:SBVtype2disparity} together with the plain link-weight threshold as were the figures in \cite{SBV}.
\subsubsection{Disparity filter--$OR$ rule with significance level $\alpha=0.01$}
Employing the $OR$ rule on the migration links with 
a significance level of $\alpha=0.01$, the number of flows (edges) passing the test was 32,294 and the number of strong components in the associated candidate multiscale backbone was 67, with the backbone having $59.0179\% $ of the total edge weights 
(that is, the total number of migrants--47,240,477--recorded in the raw data table), a "respectable" percentage. There was one giant component with 3,040 counties (cf. \cite{mendes,barbosa}), 65 isolated counties and one pair, Lipscomb and Ochiltree Counties, Texas 
(previously encountered with the two-stage algorithm). Again, the isolated (singleton) counties (none with in- or out-degree exceeding 115) were inland ones, not particularly notable as migration origins or destinations. 
\subsubsection{Disparity filter--$OR$ rule with significance levels $\alpha=0.14$ and $0.13$}
With the $OR$ rule and a much weaker significance level, $\alpha=0.14$, there are 83,693 accepted edges, and all nodes do now lie in one strong component, and $73.0026\% $ of edge weights is included. (We note that  
$83,693 > 25,329$, the number of edges needed in the two-stage analysis \cite{SlaterDendrogram}. For 
$\alpha=0.13$, there are 80,203 accepted edges and {\it two} strong components, with sparsely-populated [belittled?] King County, Texas, serving as a singleton.)
\subsubsection{Disparity filter--$AND$ rule with significance level $\alpha=0.05$}
The use of the disparity filter, using a significance level 
$\alpha=0.05$ on the $3,107 \times 3,107$ raw migration table, together with an $AND$ rule 
(that is, a link must pass the significance test, viewed as both an inflow and an outflow) yielded 25,351 links--extremely close to the 25,329 links needed to complete the  SCHC. However, with the slightly larger number of links obtained with the disparity filter, there were 181 distinct strong components (as opposed to only one with the application of the two-stage procedure). Of them, 174 were simply isolated nodes, and one "giant" one consisting of 2,836 counties. This left
six doublets (each pair comprised of contiguous counties): (1) the Georgia counties of Lincoln and Wilkes; (2) the Georgia counties of Stewart and Webster; (3) the California counties of Inyo and Mono; (4) the Nebraska counties of Nuckolls and Thayer; (5) the Kansas counties of Phillips and Smith; and (6) the Texas counties of Ochiltree and Lipscomb. (The greatest in- or out-degree--that is, the number of other counties to which migrants were sent or received--for any of these twelve counties was 146. With the exception of the first and fourth pairs listed, the same doublets were obtained in the two-stage analysis \cite{SlaterDendrogram}.) {\it All} of the 174 isolated counties were located away from the Atlantic and Pacific coasts, with only one from 
Florida and none from California or Arizona. (The greatest in- or out-degree for any of these 174 counties was 132.) So, there does not seem to be any
"Sunbelt" or "cosmopolitan" effect at work here.
\subsubsection{Disparity filter--$AND$ rule with significance level $\alpha=0.001$}
Using the $AND$ rule with $\alpha=0.001$, the resultant backbone has 
10,153 links and 525 strong components, and $42.4254\%$ of the total edge weights. The largest component consists of 2,045 
counties, while the next two largest are formed by 17 counties of the state of Montana, and 6 contiguous counties of eastern Nebraska. There are also 
two quintets (one comprised of Mississippi counties, and one of Kansas and 
Oklahoma counties) and four quartets (formed by Arkansas, Georgia, North Carolina and Texas counties).
\section{Discussion of Related Issues} \label{RelatedIssues}
\subsection{Asymmetries}
It would seem of interest and relevance to the study here of comparative properties of the two filtering procedures to also
address  a number of points
raised in the response of SBV \cite{SBV2} to the letter \cite{PBSPNAS}, commenting on their original (disparity filter) study \cite{SBV}. SBV remark that the two-state
algorithm can generate "spurious asymmetries when the original network
is symmetric." (One can initiate the iterative proportional fitting [Sinkhorn-Knopp] procedure used to convert the flow 
matrix to bistochastic form 
by first normalizing rows or by first 
normalizing columns. However, 
this should not introduce significant asymmetries in the end result if  suitable convergence is obtained.)
\subsection{Global/local issues}
The use of "globally" in the statement in 
\cite{SBV2} that "individual weights in the original matrix are globally normalized  so that they can be compared on an  equal footing" appears to suggest that the original flow matrix is simply scaled by a single number in the bistochastization--which is certainly not the case. (SBV assert that their methodology is more "local" in nature than the two-stage procedure.) In this regard, let us observe that if one doubles, say, the entries in a single row or column of the flow matrix, then the results of both the two-stage algorithm and the disparity filter are completely invariant. 

However, it is true that the decision whether or not to admit
the $ij$-link into the network backbone depends only upon the entries
in the $i$-th row and/or $j$-th column in the disparity filter, while this is certainly not the case with the bistochastic filter.

\subsection{Minimal spanning trees}
In \cite[p. 6484]{SBV}, SBV assert that "reduced networks obtained [using the minimal spanning tree (MST)]  are overly structural simplifications that destroy local cycles, clustering coefficient[s], and the clustering hierarchies often present in real world networks". The MST is the basis for the method of single-linkage clustering. Further, the strong component hierarchical clustering (SCHC) procedure \cite{tarjan2}, we have widely applied (serving as the second-stage of the two-stage algorithm), can be viewed as the extension of single-linkage clustering to weighted, directed graphs. Therefore, the remark of SBV could be thought also to extend there. However, we think that this general criticism is quite easily and naturally addressed, if one supplements the specific links in the MST, using {\it all} those links having greater weight than the minimal one employed in the MST, rather than simply those links present in the MST. (If the insertion of a link does not succeed in joining hitherto disconnected components, it is not included in the MST, no matter how large its value. Tumminello {\it et al} have suggested extracting a subnetwork that can be embedded on a surface of genus $k$, rather than a tree \cite{tumminello}. Garas and Argyrakis study an "Overlapping Tree Network", which is based on the extraction of the MST, but allows for the filtered network to have loops 
\cite{garas}.)

One additional, interesting related observation to make is that with an undirected graph, 
and in the absence of links with precisely equal values, construction of the MST always unites just two connected components. In the strong component/directed graph analogue of the MST, the insertion of a new link can, in fact,  even in the absence of links with identically equal values, join more than two strongly connected components."
\subsection{Null models}
SBV state that "there is not a clear proposal for a suitable null model to measure the statistical significance of the results" [of the two-stage 
algorithm] \cite{SBV2}. (For corroboration, they refer to an early 
1976 article \cite{japan}. However, later, in \cite{SlaterDendrogram}, we did apply a graph-theoretic isolation criterion, we had also employed in 1981 and 1983, as well 
\cite{county,qq}, to rank clusters [regions] in terms of their statistical significance (cf. \cite[sec. IV.3]{mejnewman}).) 

By way of illustration, in the 1965-70 US 3,140-county migration study, a statistical 
test of Ling \cite{ling}  (designed 
for {\it undirected} graphs), based on the difference in
the ranks of two edges, was employed in a heuristic manner
\cite[pp. 7-8]{county}. 
For example, the 3,148th largest doubly-stochastic value, 0.12972 
(corresponding to the flow from Maui County to Hawaii County), {\it united}
the four counties of the state of Hawaii. The (considerably weaker) 
7,939th largest value, 
0.07340 (the link from Kauai County, Hawaii, to Nome, Alaska), {\it integrated}
the four-county 
state of Hawaii into a much larger 
2,464-county cluster. The difference of these
two ranks, 4,192 =  7,340 - 3,148, is a measure of isolation (``survival
time'') of this state as a cluster. Reference to Table 1 in \cite{county} 
showed the significance of the state of 
Hawaii as a functional 
internal migration unit at the 0.01 level \cite[p. 7]{county}. 
(In the computation of this table, the approximation was used that
the number of edges in the relevant 
digraphs was a negligible proportion of all 
possible $3,140 \times 3,139$ edges.)

In these regards, it would also seem natural to investigate exploiting  the notion of a "random doubly-stochastic matrix" \cite{griffiths,ZKSS}. (Potentially useful then would be the seminal result 
of Birkhoff that any $n \times n$ doubly-stochastic matrix can be written as a convex combination of at most $n^2$ {\it permutation} matrices--those with a single 1 in each row and column, and zeros elsewhere.)
\subsection{Properties of bistochastic matrices}
Let us further make the general observations that powers of
bistochastic matrices are also bistochastic. Higher-order powers (in line with properties of Markov chains and the famous Perron-Frobenius Theorem), are smoother and more uniform in nature, while in the infinite limit, convergence to an 
$n \times n$ table with all entries equal to $\frac{1}{n}$ is attained. 
(This might be viewed as an interesting manifestation of all nodes being treated equally, that is, not belittled.) Mathematical 
physicists have been interested in developing conditions
that indicate when a bistochastic matrix is also {\it unistochastic} 
\cite{louck,ZKSS,unistochastic1,dita}. (This is the case if the $ij$-bistochastic entry is the square of the absolute value of the $ij$-entry of a unitary matrix.) It would be interesting to investigate whether or not unistochasticity is of value in the modeling of network flows.)
An efficient algorithm--considered as a nonlinear dynamical system--to generate 
 {\it random} bistochastic matrices has
recently been presented \cite{CSBZ} (cf. \cite{griffiths,ZKSS}). 
(Gudder has quite recently developed the concept of a bistochastic 
transition effect matrix \cite{gudder}.)
\subsection{Cosmopolitan/provincial dichotomy}
Although, by no means, have we yet systematically compared the clustering structures produced by the bistochastic and disparity filter approaches in our 1995-2000 migration analysis, the two methods--as noted in our preceding discussions--do seem to yield rather different
(but still largely contiguous) results (regions). One distinguishing, highly attractive feature of the bistochastic approach has been its ability to contrast "cosmopolitan" (hub-like or centralized) units from "provincial/local" ones. We are not aware of any comparable feature with the disparity filter.

Geographic subdivisions (or groups of subdivisions) that enter into the 
bulk of the dendrograms produced 
by the two-stage procedure at the {\it weakest} levels are those with the 
{\it broadest} ties.
These are ``cosmopolitan", hub-like areas, 
a prototypical example being 
the French capital, Paris \cite[Sec. 4.1]{tree} \cite{france}. 
Similarly, 
in parallel analyses of other 
internal migration tables, the cosmopolitan/non-provincial natures
of London \cite{siegen}, 
Barcelona \cite{spain} \cite[Sec. 6.2, Figs. 36, 37]{tree}, 
Milan \cite{gentileschi} \cite[Sec. 6.3, Figs. 39, 40]{tree} 
(cf. \cite{metron}), Amsterdam 
\cite[p. 78]{tree} \cite{masser},
West Berlin \cite[p. 80]{tree}, Moscow (the city and the oblast as a unit) 
\cite{russia} 
\cite[Sec. 5.1 and Figs. 6, 7]{tree}, Manila (coupled with suburban Rizal) 
\cite{manila}, 
Bucharest \cite{romania}, 
{\^I}le-de-Montr{\'e}al \cite[p. 87]{tree}, 
Z{\"u}rich, Santiago, Tunis and Istanbul \cite{turkish} 
were--among
others--highlighted in the respective dendrograms for their nations
 \cite[Sec. 8.2]{tree} 
\cite[pp. 181-182]{qq} \cite[p. 55]{science}. 

In our previous 
1965-70 intercounty analysis
for the US, the most cosmopolitan entities were: (1) the 
{\it centrally}-located paired
Illinois counties of Cook (Chicago) and neighboring, suburban DuPage; 
(2) the nation's capital, Washington, D. C.; and (3) the paired South
Florida (retirement) counties of Dade (Miami) and Broward (Ft. Lauderdale) 
\cite{county,partial,fields}. In general, counties with large military 
installations, large college populations or state capitals 
also interacted broadly with other areas \cite[p. 153]{county}. 
Application of the two-stage methodology to 1965-66 London inter-borough
migration \cite{masser} indicated that the three inner boroughs of Kensington
and Chelsea, Westminster, and Hammersmith acted--as a unit--in a 
cosmopolitan manner \cite[Sec. 5.2, Fig. 10]{tree}. 
(In Sec. 8.2 and Table 16 of the anthology of results \cite{tree}, 
additional geographic units and groups of
units found to be cosmopolitan with regard to migration, are enumerated.)
\subsection{Clusters/regions obtained in two-stage internal migration analyses} \label{clusters}
Geographically isolated (insular) areas--such as the Japanese islands of 
Kyushu and Shikoku \cite{japan}--emerged 
as well-defined {\it clusters} (regions)
of their constituent (seven and four, respectively) 
subdivisions (``prefectures'' in the Japanese case) 
in the dendrograms for the corresponding two-stage analyses, and similarly
 the Italian islands of Sicily and Sardinia 
\cite{gentileschi}, the North and South Islands of New Zealand, and the
Canadian islands of 
Newfoundland and Prince Edward Island \cite[p. 90]{tree}  
(cf. \cite{e,multiterminal}).
The eight counties of Connecticut, and other New England groupings, as  
further examples,  to be elaborated upon below, were
also very prominent in the highly disaggregated U. S. analysis \cite{county}. 
Relatedly, in a study based solely upon 
the 1968 movement of {\it college students} among
the fifty states, the six New England states were strongly clustered 
\cite[Fig. 1]{college}. Employing a 1963 Spanish interprovincial migration
table, well-defined regions were formed by the two provinces of
the Canary Islands, and the four provinces of Galicia \cite{spain} 
\cite[Sec. 6.2.1, Fig. 37]{tree} (cf. \cite{multiterminal}). 
The southernmost Indian states of Kerala and Madras (now Tamil Nadu) 
were strongly paired on the basis of 1961 interstate flows \cite{india}.

A detailed comparison between functional migration regions found by 
the two-stage procedure and those actually 
employed for administrative, political 
purposes in the corresponding nations is given in Sec. 8.1 and Table 15 of 
\cite{tree}. (In the 1995-2000 U. S. analysis at hand,
particularly distinct large multicounty migration regions, well describable as "French Louisiana", "Northern Lower Michigan", "Northern New England", "South Jersey"...
were found \cite[Tables I, II]{SlaterDendrogram}.)
\subsubsection{Rarity of noncontiguous groupings}
In a 1989 monograph, Gawryszewski \cite{gawryszewski,gawryszewski2} attempts to regionalize--presenting numerous dendrograms--the voivodships (provinces) of Poland on the basis of (total, rural-to-urban, and urban-to-urban) internal migration in the 1952-83 period, using the two-stage algorithm.

It should be noted that it is rare  that the two-stage
methodology yields a migration region 
composed of two or more noncontiguous subregions--even though no contiguity
information, of course,  is explicitly present in the flow table 
nor provided to the algorithm (cf. \cite{loglinear,boundary}).
A notable exception--comprehensible in terms of regional disparities in 
wealth, however--to this (topological) 
rule was the uniting of the northern 
Italian region of
Piemonte--the location of industrial Turin, where Fiat is based--with 
(poor) southern regions, {\it before} joining with central regions, in an 
aggregate 18-region  1955-70 study \cite{metron} 
\cite[p. 75]{tree} (cf. \cite{gentileschi} and \cite[p. 26]{SlaterDendrogram}) .

\section{Concluding remarks} \label{Summary}
The observation that far fewer links (25,329 {\it vs.} more than 80,203 (cf. [Fig.~\ref{fig:SBVtype2disparity}])) are needed to construct a strongly connected network backbone in the bistochastic case 
\cite{SlaterDendrogram} than in the disparity analysis, appears, for the specific dataset at hand, to serve as a positive aspect of the bistochastic filtering methodology. 
Another positive feature is the fact that the correlations of the logarithms of the 735,531 nonzero raw (unadjusted) intercounty migration flows with the logs of the corresponding bistochasticized links is 0.318379, but with the disparity filter significance levels, considerably stronger in nature, that is,  -0.849757 (using the $AND$ rule) and -0.886816 ($OR$ rule). Thus, in terms of the criteria put forth by Serrano, Bogu{\~n}{\'a} and Vespignani themselves \cite{SBV}, it appears that the bistochastic filter, desirably, "belittles" small flows (and nodes) less so than does the SBV disparity filter.

Nevertheless,more detailed comparative studies, in particular as to "topological" properties, as called for by 
SBV, are certainly in order.
(One important topological property is that of
strong connectedness, upon which the second stage of the two-stage algorithm \cite{tree} is based. Another is the pronounced rarity--even in the absence of imposition of contiguity constraints--of 
noncontiguous couplings observed [sec.~\ref{clusters}] in the numerous two-stage internal
migration analyses of various nations that have been conducted.
This phenomenon is well exhibited in the 3,000+ U. S. county dendrogram in the supplementary material, and would be immediately apparent in the correspondingly reordered doubly-stochastic internal migration matrix 
(Fig.~\ref{fig:ReOrdered1}) if county-state labels could be attached to the rows and columns.)
\subsection{{\it Gedanken} experiment}
Somewhat contrastingly, one might consider a {\it gedanken} experiment, in which  a ("non-multiscale") network in precisely bistochastic (or proportional thereto) form is empirically (and remarkably!) obtained.
Then, bistochastic filtering would not alter the internodal scores in any manner, while
presumably disparity filtering would. Clearly, the correlation between
the filtered and raw scores would be perfect (1.0) in the bistochastic scenario and weaker (in contrast to our main U. S. intercounty migration analysis above) with the disparity filter.
\subsection{Domains of applicability}
Finally, let us remark that the two methodologies would seem to work better in different domains: the disparity filtering approach of SBV may be more appropriate when small samples are involved and the use of measures of significance of flows are indicated, while the two-stage methodology employs (bistochastic) measures of association, and perhaps is more effective in constructing a network backbone when considerable data are at hand. This is an analytical dichotomy--that of finding a suitable balance between significance and sample size--which had already arisen 
several decades ago in the  influential "transaction flow" models of Savage, and Deutsch and Brams \cite{savage,brams} in the political science literature (cf. \cite{mosteller}).

\begin{acknowledgments}
I would like to express appreciation to the Kavli Institute for Theoretical Physics (KITP) for technical support, to Andrew Carter for a discussion on these subjects.
\end{acknowledgments}

\bibliography{MultiScaleDeNovo2}% Produces the bibliography via BibTeX.

\end{document}